\documentclass[a4paper,11pt,DIV=calc]{article}

\usepackage[a4paper,left=30mm,right=30mm,top=30mm,bottom=30mm,marginpar=23mm]{geometry} 

\usepackage{hyperref}
\usepackage{subfig}
\usepackage{amsmath,amssymb}
\usepackage{bbm}
\usepackage{amsthm}
\usepackage{pgfplots}
\usepackage{graphicx}
\usepackage{MnSymbol}
\usepackage{color}
\theoremstyle{plain} \newtheorem{Theorem}{Theorem}
\theoremstyle{plain} \newtheorem{Proposition}[Theorem]{Proposition}
\theoremstyle{definition} \newtheorem{Example}{Example}

\newcommand{\id}{{\mathbb{I}}}

\DeclareMathOperator{\Tr}{Tr}

\newcommand{\dm}{\operatorname{d_M}}

\newcommand{\m}{\mu}
\newcommand{\la}{\lambda}

\newcommand{\so}{{\rm{SO(3)}}}

\newcommand{\tp}{\otimes}

\newcommand{\bb}{\ensuremath{\mathbf{b}}}

\newcommand{\mm}{\ensuremath{\mathbf{m}}}

\newcommand{\al}{\alpha}

\newcommand\eo{\eta_1}
\newcommand\et{\eta_2}

\mathchardef\mdash="2D

\newcommand{\aaa}{\mathbf{a}}
\newcommand{\vv}{\mathbf{v}}
\newcommand{\ee}{\mathbf{e}}
\newcommand{\nn}{\mathbf{n}}
\newcommand{\dd}{\mathbf{d}}
\newcommand{\pp}{\mathbf{p}}

\usepackage{authblk}

\usepackage{fancyhdr}

\newcommand\shorttitle{The morphology of lath martensite: a new perspective}
\newcommand\authors{K. Koumatos, A. Muehlemann}

\begin{document}
%\graphicspath {{}}

\makeatletter
\def\@maketitle{%
  \newpage
  \null
  \vskip 2em%
  \begin{center}%
  \let \footnote \thanks
    {\Large\bf  \@title \par}%
    \vskip 1.5em%
    {\normalsize
      \lineskip .5em%
      \begin{tabular}[t]{c}%
        \@author
      \end{tabular}\par}%
    \vskip 1em%
    {\normalsize \@date}%
  \end{center}%
  \par
  \vskip 1.5em}
\makeatother
\title{{The morphology of lath martensite: a new perspective}}

\author{Konstantinos Koumatos%
  \thanks{\texttt{konstantinos.koumatos@gssi.infn.it}}}
\affil{\small\textit{Gran Sasso Science Institute, }\\ \textit{Viale Fransesco Crispi 7,} \\ \textit{67100, L'Aquila, Italy}}

\author{Anton Muehlemann%
  \thanks{\texttt{muehlemann@maths.ox.ac.uk}}}
\affil{\small\textit{Mathematical Institute, University of Oxford,}\\\textit{ Andrew Wiles Building, Radcliffe Observatory Quarter, Woodstock Road,} \\ \textit{Oxford OX2 6GG, United Kingdom}}

\date{Dated: \today}

\maketitle

% 
% \title{\bf{The morphology of lath martensite: a new perspective}}
% \author[1]{Konstantinos Koumatos\thanks{konstantinos.koumatos@gssi.infn.it}}
% \author[2]{Anton Muehlemann\thanks{muehlemann@maths.ox.ac.uk}}
% \affil[1]{Gran Sasso Science Institute, Viale Fransesco Crispi 7, 67100, L'Aquila, Italy}
% \affil[2]{Mathematical Institute, University of Oxford, Andrew Wiles Building, Radcliffe Observatory Quarter, Woodstock Road, Oxford OX2 6GG, United Kingdom}
% 
% \renewcommand\Authands{ and }
% 
% % \author{Konstantinos Koumatos}
% % %\address{\textit{Konstantinos Koumatos:} Gran Sasso Science Institute, Viale Fransesco Crispi 7, 67100, L'Aquila, Italy}
% % %\email{konstantinos.koumatos@gssi.infn.it}
% % 
% % \author{Anton Muehlemann}
% % %\address{\textit{Anton Muehlemann:} Mathematical Institute, University of Oxford, Andrew Wiles Building, Radcliffe Observatory Quarter, Woodstock Road, Oxford OX2 6GG, United Kingdom}
% % %\email{muehlemann@maths.ox.ac.uk}
% 
% 
 \hypersetup{
  pdfauthor = {Konstantinos Koumatos (Gran Sasso Science Institute) and Anton Muehlemann (University of Oxford)},
  pdftitle = {The morphology of lath martensite: a new perspective},
  pdfsubject = {MSC (2010): 74N05, 74N15},
  pdfkeywords = {lath martensite, microstructure, twins within twins, (557) habit planes, double shear theories, energy minimisation, non-classical interfaces}
}
% % 
% \maketitle

\begin{abstract}
A mathematical framework is proposed to predict the features of the (5\,5\,7) lath transformation in low-carbon steels based on energy minimisation. This theory generates a one-parameter family of possible habit planes and a selection mechanism then identifies the (5\,5\,7) normals as those arising from a deformation with small atomic movement and maximal compatibility. While the calculations bear some resemblance to those of double shear theories, the assumptions and conclusions are different. Interestingly, the predicted microstructure morphology resembles that of plate martensite, in the sense that a type of twinning mechanism is involved.
\vspace{4pt}



\vspace{4pt}
\end{abstract}

\section{Introduction}
\label{intro}
The present article proposes a theory that predicts the formation of habit plane normals very close to $(5\,5\,7)$, observed in steels with low carbon content (less than $0.4\%$ \cite{Wayman}). In fact, \emph{all} the predicted habit plane normals are almost exactly\footnote{E.g. one of the predicted normals is $\mm=(0.4813,0.4956,0.7228)$ with $\angle(\mm,(2\,2\,3))=0.7^\circ$ and \\ $\|\mm-(2\,2\,3)\|_2= 0.012$, where $\|\cdot\|_2$ denotes the Euclidean distance.} $(2\,2\,3)$. Widely accepted models that result in $(5\,5\,7)$ habit planes are double shear theories, e.g. \cite{Acton, Ross}, and some of the most accurate explanations are due to the algorithm developed by Kelly \cite{Kelly}. These can be seen as generalisations of the so-called phenomenological theory of martensite most notably developed by Wechsler, Liebermann \& Read \cite{Read} to explain the $(3\,10\,15)$ habit planes in plate martensite and Bowles \& MacKenzie \cite{BM123} who applied their theory to explain the $(2\,5\,9)$ and $(2\,2\,5)$ 
habit planes also in plate 
martensite.

A short-coming of single/double shear theories is the lack of a selection mechanism that picks the right lattice invariant shearing systems (see e.g.~\cite[Table 1]{Kelly}), in turn leading to a large number of input parameters. To overcome this, one approach is to only allow shearing systems that arise from mechanical twinning, cf.~\eqref{Fl}. Indeed, in the context of single shear theories \cite{Read} and \cite{BM3} made this assumption and proposed that the martensite plates consist of a \emph{``stack of twin-related laths''}. As pointed out in \cite[p.~381f]{KellyBook}, TEM investigations in \cite{Kelly61} showed that even though \emph{``under the optical microscope there is little sign that this is the case, {\normalfont{[...]}} when such steels are examined with the transmission electron microscope, arrays of very thin $\{1\,1\,2\}_M$ twins are indeed found''} - marking a significant success from a theoretical prediction to an observed feature in plate martensite. 

Regarding $(5\,5\,7)$ habit planes in lath martensite, it can be shown that for any reasonable choice of lattice parameters (see also Figure~\ref{FigSimplNormals}), a single shear theory with shearing systems arising from twinning in bcc crystals cannot give rise to them. However, in this paper we show that by introducing another level of twinning (``twins within twins'') we are not only able to explain $\{5\,5\,7\}$ habit planes but also predict them by showing that it is the only possible family of habit planes that satisfies a condition of maximal compatibility and a condition of small overall atomic movement. Under this interpretation each lath may be seen as a region of twins within twins. In other materials, twins within twins have commonly been observed purely in martensite \cite{Arlt,Denquin} as well as along interfaces with austenite \cite{BKSicomat08}. Moreover for lath martensite it has been observed in \cite{Sandvik} that \emph{``Twinning within a lath may be heavy {\normalfont{[...]}}. In any 
event, whenever an 
exact twin relationship was identified, it was found to be a result of twinning within a
given lath and not of a twin relation between adjacent laths. {\normalfont{[...]}} It is believed that the existence
of heavily twinned local regions of laths, which may appear as separate
laths in contrast images, may have caused some misinterpretation in earlier work on lath martensite.''}

As in single shear theories, twinning of twins is macroscopically equivalent to a simple shear of a simply twinned system (cf. text below \eqref{Flm}) and thus the step from twins within twins is in analogy with the step from a single to a double shear theory. %We wish to stress that the predictions made in this article concern macroscopic quantities, e.g. average strains or habit planes of lath boundaries, and thus we do not propose any definite morphology for the internal structure of laths. In particular the predicted macroscopic quantities can either be seen as having arisen from twinning, slip or a combination of the two.     

The strength of the theory presented here is that it enables one to predict $(5\,5\,7)$ habit planes only assuming the lattice parameters of austenite and martensite. This is particularly striking when compared to the double shear theory in \cite{Kelly} where the \emph{``calculation strategy was to select one of the possible $S_2$ {\normalfont{[second shear]}} systems and then perform calculations for the $S_1$ {\normalfont{[first shear]}} systems {\normalfont{[...]}} over a range of values of $g_2$ {\normalfont{[the shearing magnitude of $S_2$]}}. {\normalfont{[...]}} The sign of $g_2$ was selected by trial and error depending on whether the habit plane moved towards or away from $(5\,5\,7)$.''} 
\section{A model of phase-transformations based on nonlinear elasticity} \label{SecModel}
The theory proposed in this article is derived from the Ball-James model \cite{BallFine} - based on nonlinear elasticity and energy minimisation - and expands on previous work by Ball \& Carstensen \cite{BallNon} on the possibility of nonclassical austenite-martensite interfaces. Even though lath martensite is commonly associated with a high dislocation density, slip and plasticity the present model does not take such effects into account (see also Section \ref{SecConcl}). 
%the transformation from austenite to martensite must start elastically. Within this elastic regime the material is free to choose any morphology and corresponding macroscopic strain that minimises energy. Once the yield point is reached the material deforms plastically to release the excess stress. However, we take the viewpoint that this plastic deformation will only change the material's internal structure in order to accommodate the macroscopic strain already reached elastically.

Interestingly, the Ball-James model recovers the results of the phenomenological theory of martensite, as can be seen through a comparison of the derived formulae for the habit planes between twinned martensite and austenite (cf. \cite[eq. (5.89)]{BallFine} and \cite[eq. (33)-(34)]{Read}). For a self-contained account of the phenomenological theory or of the Ball-James model the reader is referred to the monographs \cite{Bhadeshia} and \cite{Bha}, respectively, both addressed to non-specialists.

In the Ball-James model, which neglects interfacial energy, microstructures are identified through minimising sequences $y^k$, $k=1,2,\dots$, for a total free energy of the form
\begin{equation}
 E(y):= \int_\Omega W_\theta(Dy(x)) dx. \label{FreeEnergy} \tag{E}
\end{equation}
Here, $\Omega$ is a region representing the reference configuration of undistorted austenite at the transformation temperature and $y(x)$ denotes the deformed position of particle $x$ in $\Omega$. We remark that passing to the limit in these minimising sequences, corresponds in a very precise way to passing from a micro- to a macroscale, so that the limits themselves can be identified with the macroscopic deformations. The energy density $W_\theta(F)$ depends only on the deformation gradient $F=Dy$, a 3$\times$3 matrix with positive determinant, and the temperature $\theta$. Also $W_\theta$ is assumed frame indifferent, i.e. $W_\theta(RF)=W_\theta(F)$ for all rotations $R$ - that is, for all 3$\times$3 matrices in $\so = \{R: R^TR =\id,\, \det R=1\}$ and must respect the symmetry of the austenite, i.e. $W_\theta(FP)=W_\theta(F)$ for all rotations $P$ leaving the austenite lattice invariant. For cubic austenite there are precisely 24 such rotations. Below the transformation temperature, $W_\theta$ is 
minimised 
on the set $K$ of martensitic energy wells, that is $W_\theta(F)$ is minimal for $F\in K = \bigcup_{i=1}^N \so U_i$. The $3\times 3$, positive-definite, symmetric matrices $U_i$ are the pure stretch components of the transformation strains mapping the parent to the product lattice. For example, in the case of fcc to bcc or fcc to bct, these are given by the three Bain strains %\comr{revisit}
\begin{align*}
U_1= B_1 = \operatorname{diag}(\eta_2,\eta_1,\eta_1),\quad U_2= B_2 =\operatorname{diag}(\eta_1,\eta_2,\eta_1), \quad
U_3= B_3 =\operatorname{diag}(\eta_1,\eta_1,\eta_2),
\end{align*}
where $\eta_1=\frac{\sqrt{2}a}{a_0}$ and $\eta_2=\frac{c}{a_0}$. Here $a_0$ is the lattice parameter of the fcc austenite and $a$, $c$ are the lattice parameters of the bct martensite ($a=c$ for bcc). The notation $B_1$, $B_2$, $B_3$ has been chosen to emphasise that we are in the Bain setting and to stay consistent with the literature. We remark that the Bain transformation \cite{Bain} is widely accepted as the transformation from fcc to bct/bcc requiring least atomic movement; for a rigorous justification see \cite{BainPaper}.

A convenient way to understand the relation between microstructures and minimising sequences is illustrated by the following example (cf. \cite{BallFine}).
\begin{Example}{(Austenite-twinned martensite interface)}\label{example1}\\
\begin{figure}[ht]
	\centering
	\def\svgwidth{0.46\columnwidth}
	\begingroup
    \setlength{\unitlength}{\svgwidth}
  \begin{picture}(1,0.5)%
     \put(0,0){\includegraphics[width=\unitlength]{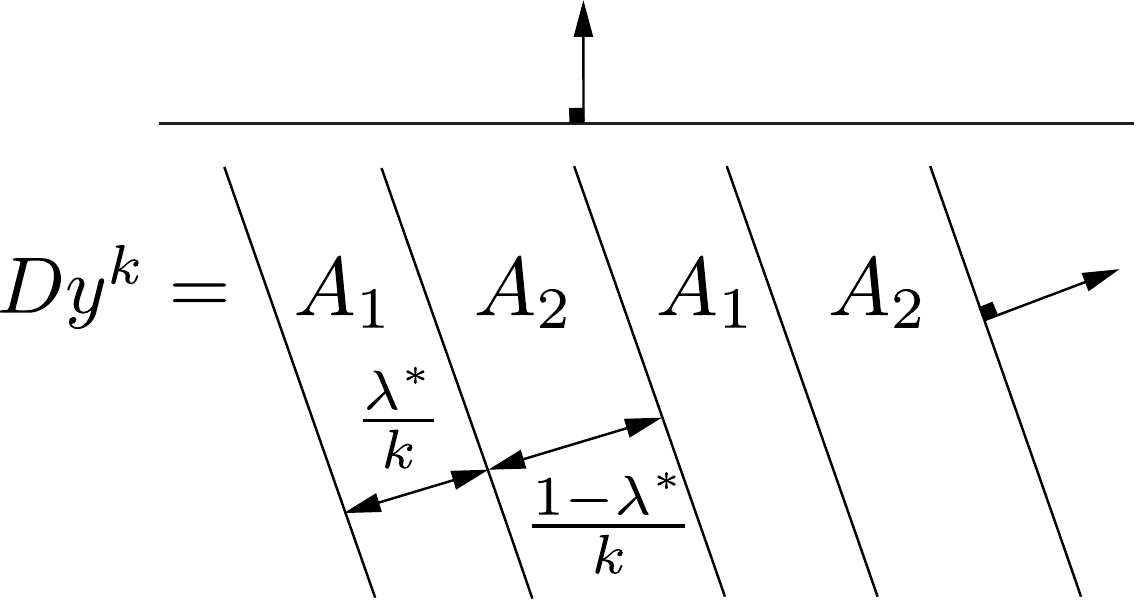}}%
     \put(0.525,0.44){\color[rgb]{0,0,0}\makebox(0,0)[lb]{\smash{$\mm$}}}
     \put(.88,0.284){\color[rgb]{0,0,0}\makebox(0,0)[lb]{\smash{$\nn$}}}
  \end{picture}%
\endgroup
	\caption{An energy minimising sequence modelling twinned martensite. The limit $k\to\infty$ corresponds to an IPS leaving the plane with normal $\mm$ invariant.}
	\label{fig:laminate}
\end{figure}
Suppose that a region of martensite is occupied by an array of twin related variants $A_1$~and~$A_2$ with relative volume fractions $1-\lambda$ and $\lambda$. The strains $A_1$ and $A_2$ in general cannot be invariant plane strains (IPS), equivalently, they cannot form a fully coherent interface with austenite, represented in this model by the identity matrix $\id$. However, for specific volume fractions $\lambda^\ast$ (given by \eqref{Eqmu} for the Bain strain), the average deformation strain of the twinned region $(1-\lambda^\ast) A_1 +\lambda^\ast A_2$ may indeed become an IPS. In terms of the nonlinear elasticity model, this inability to form a fully coherent interface at the microscopic level, implies that the austenite-twinned martensite configuration cannot exactly minimise the energy \eqref{FreeEnergy}. Nevertheless, one can construct an energy minimising sequence $y^k$, $k=1,2,\dots$, with gradients $Dy^k$ as in Figure~\ref{fig:laminate}. The limit of this sequence is precisely the average strain/total shape deformation $(1-\lambda^\ast) A_1 + \lambda^\ast A_2$ that is an IPS. Although this average strain does not minimise the energy \eqref{FreeEnergy}, it can be interpreted as a minimiser of a corresponding macroscopic energy. At the microscopic level, one would observe some specific element of the above minimising sequence rather than the limit, due to having neglected interfacial energy.
\end{Example}
The set of all matrices that can arise as limits of minimising sequences for the energy \eqref{FreeEnergy} is referred to as the quasiconvex hull of the martensitic energy wells $K$, denoted by $K^{qc}$. Hence, the set $K^{qc}$ corresponds to all possible homogeneous total shape deformations that are energy minimising at the macroscopic level. Then, the requirement that (up to an overall rotation $R$) a martensitic microstructure with a total shape deformation $F\in K^{qc}$ is a strain leaving the plane with normal $\mm$ invariant, amounts to finding a vector $\bb$ such that
\begin{equation}\label{eq:IPS}
RF = \id +\bb\otimes \mm \in K^{qc}.
\end{equation}
Here, $\bb\otimes \mm$ denotes the 3$\times$3 matrix $(\bb\otimes \mm)_{ij}= b_im_j$. Writing $\hat \bb= \bb/|\bb|$, yields the equivalent expression
$RF = \id + |\bb| \hat \bb \tp \mm$,
implying that $RF$ is the IPS given by a shear on the plane with normal $\mm$, with shearing direction $\hat \bb$ and shearing magnitude $|\bb|$. In particular, if the transformation from parent to product phase is volume-preserving, $F$ is a simple shear, corresponding to the vectors $\bb$ and $\mm$ being perpendicular. We note that \eqref{eq:IPS} equivalently says that $RF$ can form a fully coherent planar interface with austenite of normal $\mm$. Also by frame-indifference, the austenite can be represented by any rotation. Hence, a further rotation $Q$ of the martensite results in $QRF=Q+(Q\bb)\tp\mm$, so that the rotated martensite can still form a fully coherent interface with austenite of the same normal $\mm$. That is, rotations from the left cannot change the habit plane normal.
\section{Comparison to the phenomenological theory}
A common feature in both the phenomenological theory and the Ball-James model is to construct (up to an overall rotation $R$) a total shape deformation $F$ that is an IPS. In the literature, various algorithms have been proposed for the calculation of the corresponding elements of the shear, i.e. the magnitude, direction and normal (cf. \eqref{eq:IPS}). For example, see \cite{BM2,Read} in the context of twinning/single shear theories and \cite{Acton,Ross} for double shear theories.

In general, the problem of finding an overall rotation $R$ and shearing elements such that $RF=\id+\bb\otimes \mm$ can be simplified by only considering the Cauchy-Green strain tensor $C=(RF)^T(RF)=F^TF$ and thus factoring out the overall rotation $R$. The following Proposition (\cite[Proposition 4]{BallFine}), allows one to calculate the shearing elements $\bb$ and $\mm$ in terms of the principal stretches and stretch vectors of $F$. The overall rotation $R$ can then be found by substituting $\bb$ and $\mm$ back into equation \eqref{eq:IPS}.
\begin{Proposition}\label{PropBJ}
 Let $C\neq \id$ be a symmetric $3\times 3$ matrix with ordered eigenvalues $\lambda_1 \leq \lambda_2 \leq \lambda_3$. Then $C$ can be written as
 \begin{equation*} 
  C=(\id+\mm \tp \bb)(\id+\bb \tp \mm)  
 \end{equation*}
for some $\bb, \mm$ if and only if $\lambda_1 \geq 0$ and $\lambda_2=1$.
Then, there are at most two solutions given by
 \begin{align*}
  &\bb=\frac{\rho}{\sqrt{\lambda_1^{-1}-\lambda_3^{-1}}}\left( \sqrt{\lambda_1^{-1}-1}{\bf v}_1+\kappa \sqrt{1-\lambda_3^{-1}} {\bf v}_3 \right) ,  \\
  &\mm=\rho^{-1} \left( \frac{\sqrt{\lambda_3}-\sqrt{\lambda_1}}{\sqrt{\lambda_3-\lambda_1}}\right) (- \sqrt{1-\lambda_1}{\bf v}_1+\kappa \sqrt{\lambda_3-1}{\bf v}_3),  
 \end{align*}
 where $\rho \neq 0$ is a normalisation constant, $\kappa \in \lbrace -1,1\rbrace$ and ${\bf v}_1,{\bf v}_3$ are the (normalised) eigenvectors of $C$ corresponding to $\lambda_1$ and $\lambda_3$.
\end{Proposition}

\subsection*{A short interlude on martensite twins}
In the material science literature twins are often described as two phases related by a specific $180^\circ$ degree rotation or, equivalently, a reflection. In the mathematical literature a twin is usually characterised by the existence of a rank-one connection between the two deformation strains $A_1$, $A_2$ corresponding to the two phases, i.e. the existence of vectors $\aaa$ and $\nn$ such that
\[ A_2 = A_1 + \aaa \otimes \nn.\] 
A fully coherent interface between the two phases is then given by the plane of normal $\nn$. This is because for any vector $\vv$ on that plane, i.e. $\vv\cdot\nn = 0$, we obtain $A_2\vv = A_1\vv + (\vv\cdot\nn) \aaa = A_1\vv$. Also, note that $A_2= (\id + \aaa\tp A_1^{-T}\nn)A_1$ so that the lattice on the one side of the interface can be obtained by shearing the lattice on the other side along the twin plane $\frac{A_1^{-T}\nn}{|A_1^{-T}\nn|}$, in the shearing direction $\frac{\aaa}{|\aaa|}$ with shearing magnitude $|\aaa||A_1^{-T}\nn|$. The latter expression enables one to calculate the vectors $\aaa$ and $\nn$ by Proposition~\ref{PropBJ} through the identification $F=A_2A_1^{-1}$, that is the relative deformation between the two phases is an IPS. In view of single shear theories, the above expression can equivalently be written as $A_2=A_1(\id + A_1^{-1}\aaa\tp\nn)$ and thus $A_2$ can be obtained as a shear of the parent lattice, followed by $A_1$.

Hence, in the case of twins between two martensitic energy wells $\so U_i$ and $\so U_j$ one needs to solve the equation 
\begin{equation}\label{eq:twin}
QU_j=U_i+\aaa\tp\nn
\end{equation}
for the rotation matrix $Q$ and the twinning elements $\aaa$ and $\nn$. If the transformation strains $U_i$ and $U_j$ are related by a $180^\circ$ rotation, this calculation simplifies significantly by Mallard's Law (see \cite[Result 5.2]{Bha} or below). In particular, this assumption holds for $U_i=B_i$ and $U_j=B_j$, i.e. for the Bain transformation from fcc to bct/bcc.
\begin{Proposition}{(Mallard's Law)}\label{LemmaMal}\\
 Let $U$ and $V$ satisfy $V=PUP$ for some $180^\circ$ rotation $P$  about a unit vector $\ee$, i.e. $P=-\id + 2 \ee \tp \ee$. Then the equation $QV = U + \aaa \tp \nn$ admits two solutions given by
\begin{align} \tag{I}\label{EqM1}
  \aaa&=2 \left(\frac{U^{-T}\ee}{|U^{-T}\ee|^2}-U\ee\right), \ &&\nn=\ee,\\ 
  \aaa&=\frac{2N}{|U\ee|^2}U\ee, \ &&\nn=\frac{1}{N}\left(|U\ee|^2 \ee -U^TU\ee\right). \tag{II} \label{EqM2}
\end{align}
In each case, $Q= (U+ \aaa \tp \nn)V^{-1}$.
\end{Proposition}
\noindent We conclude this interlude by remarking that twins described by the first solution in Mallard's law are \emph{Type I} twins and the corresponding lattices are related by a 180$^\circ$ rotation about the twin plane $\frac{U^{-T}\nn}{|U^{-T}\nn|}$. The second solution in Mallard's law describes \emph{Type II} twins and the lattices are related by a 180$^\circ$ rotation about the shearing direction $\frac{\aaa}{|\aaa|}$. It may happen, and it does for the Bain strains, that there are two rotations by 180$^\circ$ relating $U$ and $V$. In this case, there are seemingly four solutions from Mallard's Law, however, Proposition~\ref{PropBJ} says that there cannot be more than two. Indeed, the Type I solution using one 180$^\circ$ rotation is the same as the Type II solution using the other 180$^\circ$ rotation and vice versa. In particular, the lattices on either side of the interface are related by both a 180$^\circ$ rotation about the twin plane and a 180$^\circ$ rotation about the shearing direction. 
Solutions of 
this type are \emph{compound} twins. 

\subsection{Single shear theories}
Henceforth, we only consider the Bain strains $B_1$, $B_2$ and $B_3$ for the fcc to bct/bcc transformation in steel. In single shear theories the total shape deformation is assumed to be decomposable into $F=RBS$ where $R$ is a rotation, $B$ is one of the Bain strains and $S=\id+ \dd \tp \pp$ is a shear whose specific form varies in the literature. In the Ball-James theory, the total shape deformation $F$ must be macroscopically energy minimising, thus restricting the form of the shear $S$. The most important case is when $S$ arises from twinning. As in Example~\ref{example1}, the average strain corresponding to a twinning system between $A_1=B_1$ and $A_2=QB_2$, satisfying \eqref{eq:twin}, with volume fractions $1-\lambda$ and $\lambda$, respectively, is given by $(1-\lambda) B_1+\lambda QB_2=B_1+\lambda \aaa \tp \nn$, for any $\lambda\in (0,1)$, where the elements $Q$, $\aaa$ and $\nn$ can be calculated by Mallard's Law (cf. Proposition~\ref{LemmaMal}) applied to $U = B_1$ with either $\ee=(1,1,0)
$ or $\ee=(1,
-1,0)$, i.e. the resulting twins are compound. By simple algebraic manipulation, the average strain can be written as
\begin{equation}
B_1(\id+\lambda B_1^{-1}\aaa \tp \nn)=B_1S_\lambda, \label{Fl}
\end{equation}
i.e. a single shear $S_\lambda=\id+\lambda B_1^{-1}\aaa \tp \nn$, with $\det S_\la=1$, of the parent lattice followed by the Bain strain $B_1$. 

\begin{figure}[h]
  \centering
  {\includegraphics[width=0.65\textwidth]{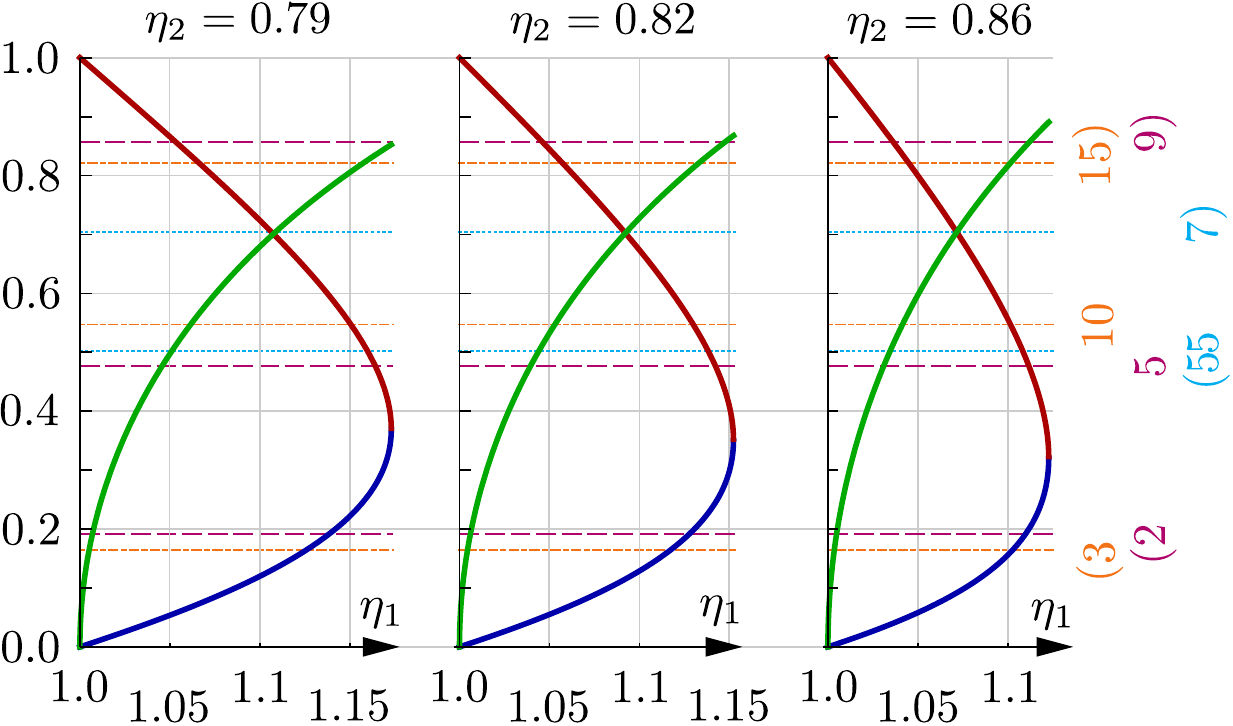}} 
  \caption{Coordinates of habit plane normal (blue,\,red,\,green) arising from simple twinning for different values of $\eo, \et$.} 
  \label{FigSimplNormals}
\end{figure}

By Proposition~\ref{PropBJ}, to make the total shape deformation $F_\la = RB_1S_\la$ an IPS, the volume fraction $\la$ needs to be chosen such that the middle eigenvalue of $F^T_\la F_\la$ is equal to one. In particular, since one of the eigenvalues must be made equal to one, the expression $\det(F^T_\la F_\la-\id)$ must vanish, giving rise to the two solutions $\lambda^\ast$ and $1-\lambda^\ast$, where
\begin{equation} \label{Eqmu}
 \la^\ast=\frac{1}{2} - \frac{1}{2}\frac{1}{\et^2-\eo^2}\sqrt{(2-\et^2-\eo^2)(\eo^2-2\eo^2\et^2+\et^2)}.
\end{equation}
It is important to then check that it is indeed the middle eigenvalue that is equal to one. For each of the values $\la^\ast$, $1-\la^\ast$, we can calculate two habit plane normals according to Proposition~\ref{PropBJ}. One of these normals is, up to normalisation, given by $(h\,k\,1)$, where
 \begin{align*}
  h=\frac{1}{2\sqrt{\eo^2-1}}\left(\sqrt{{\eo^2+\et^2-2\eo^2\et^2}}-\sqrt{{2-\eo^2-\et^2}}\right),\\
  k=\frac{1}{2\sqrt{\eo^2-1}}\left(\sqrt{{\eo^2+\et^2-2\eo^2\et^2}}+\sqrt{{2-\eo^2-\et^2}}\right).
 \end{align*}
By considering the remaining normals and all possible pairs of twin related Bain strains, one recovers the entire family of normals $\{h\,k\,1\}$. In Figure~\ref{FigSimplNormals}, the components of one of these normals are plotted for a typical range of lattice parameters $\eo$, $\et$ in the fcc to bct/bcc transformation. We immediately note that for $\eo=1.1$ and $\et=0.86$ the predicted habit plane normal arising from simple twinning is almost exactly $(3 \, 10 \, 15)$ and for $\eo=1.11$ and $\et=0.86$ the habit plane normal is almost exactly $(2 \, 5 \, 9)$. The corresponding ratios of tetragonality are given by $c/a=\sqrt{2}\eo / \et \approx 1.105$ and $  \approx 1.095$ respectively. This in excellent agreement with the observations in e.g. \cite{Richards} of $(2 \, 5 \, 9)$ habit planes in steel with carbon content in the range $1.4-1.8$ wt-$\%$ as well as the theoretical and experimental results in e.g. \cite{Kelly61}, \cite{Wata} and \cite{Gren} of $(3 \, 10\,  15)$ habit planes in highly tetragonal 
martensite.  

\subsection{Double shear theories} \label{SecDoubleShear}
Similarly, in double shear theories the total shape deformation is assumed to be decomposable into $F=RBS_2S_1$ where $R$ is a rotation, $B$ is one of the Bain strains and $S_1$, $S_2$ are two shears. Above we have seen how twinning can be regarded as an instance of a single shear theory. 
% \begin{figure}[ht]
%   \centering
%   {\includegraphics[width=0.37\textwidth]{DoubleLamD.eps}} 
%   \caption{A configuration of twins within twins macroscopically leaving the plane with normal $\mathbf{k}$ invariant.\comr{swap,$m_\la$}} 
%   \label{FigDouble}
% \end{figure}
% 
% 
\begin{figure}[ht]
	\centering
	\def\svgwidth{0.46\columnwidth}
	\begingroup
    \setlength{\unitlength}{\svgwidth}
  \begin{picture}(1,0.6)%
     \put(0,0){\includegraphics[width=\unitlength]{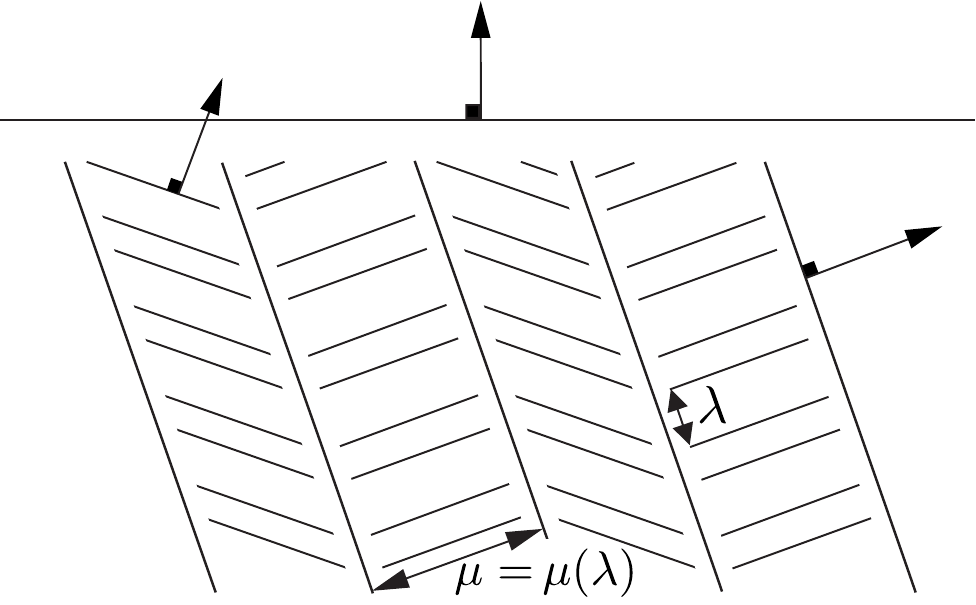}}%
     \put(0.51,0.51){\color[rgb]{0,0,0}\makebox(0,0)[lb]{\smash{$\mathbf{k}$}}}
     \put(.825,0.38){\color[rgb]{0,0,0}\makebox(0,0)[lb]{\smash{$\mm_\la$}}}
     \put(.146,0.44){\color[rgb]{0,0,0}\makebox(0,0)[lb]{\smash{$\nn$}}}
  \end{picture}%
\endgroup
	\caption{A configuration of twins within twins macroscopically leaving the plane with normal $\mathbf{k}$ invariant.}
	\label{FigDouble}
\end{figure}
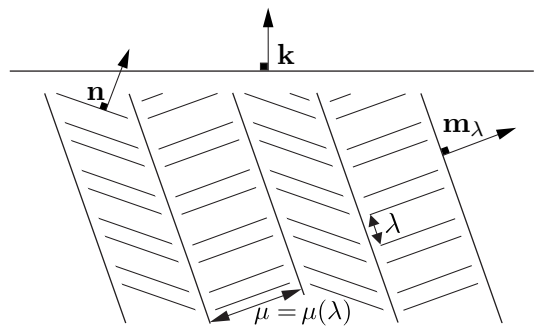
In this section we show how an additional level of twinning, i.e. twins within twins, results in an instance of a double shear theory, consistent with the Ball-James model. To visualise this type of microstructure, we revisit the construction in Example~\ref{example1} to construct \emph{two} twinning systems one with $A_1=B_1$, $A_2=QB_2$ and average strain $B_1+\la \aaa \tp \nn$ and another one with say $A_1=B_1$, $A_2=Q'B_3$ and average strain $B_1+\la \aaa' \tp \nn'$. For each $\la$ there exists a rotation $R'_\la$ such that 
\[
R'_\la(B_1+\la \aaa' \tp \nn') = (B_1+\la \aaa \tp \nn) + \bb_\la \tp \mm_\la,
\]
and thus the two twinning systems are macroscopically compatible with a fully coherent interface of normal $\mm_\la$ between them (cf. Figure~\ref{FigDouble}). The elements $R'_\la$, $\bb_\la$ and $\mm_\la$ can be calculated by Mallard's Law (cf. Proposition~\ref{LemmaMal}) applied to $U = B_1+\la \aaa \tp \nn$ with $\ee=(0,1,1)$, giving rise to a Type I and a Type II solution. Unlike in the single shear theory, these twins are not compound and therefore we distinguish these two solutions by the superscript $\al$, $\al = 1,2$ for Type I and Type II respectively. Finally, it can be shown that the volume fractions $\la$ in each of the twinned regions must necessarily coincide. As in Example~\ref{example1}, we can construct an array of twins between the two twinned regions with respective volume fractions $1-\mu$ and $\mu$ and average strain given by 
\begin{equation} \label{Flm}
B_1+\la \aaa \tp \nn + \mu \bb^\al_\la \tp \mm^\al_\la,\quad  \la,\mu \in (0,1).
\end{equation}
Simple algebraic manipulation allows one to write \eqref{Flm} as $B_1S^\al_{2}(\la,\m)S_{1}(\la)$ where $S_1\equiv S_1(\la)=\id+\la B_1^{-1}\aaa \tp \nn$ and $S_2\equiv S^\al_{2}(\la,\m)=\id+\m B_1^{-1}\bb^\al_\la \tp S_{1}^{-T}\mm^\al_\la$ and thus an instance of a double shear theory. We note that $\det S_1=\det S_2=1$.

\begin{figure}[ht]
  \centering
  \includegraphics[width=.66\columnwidth]{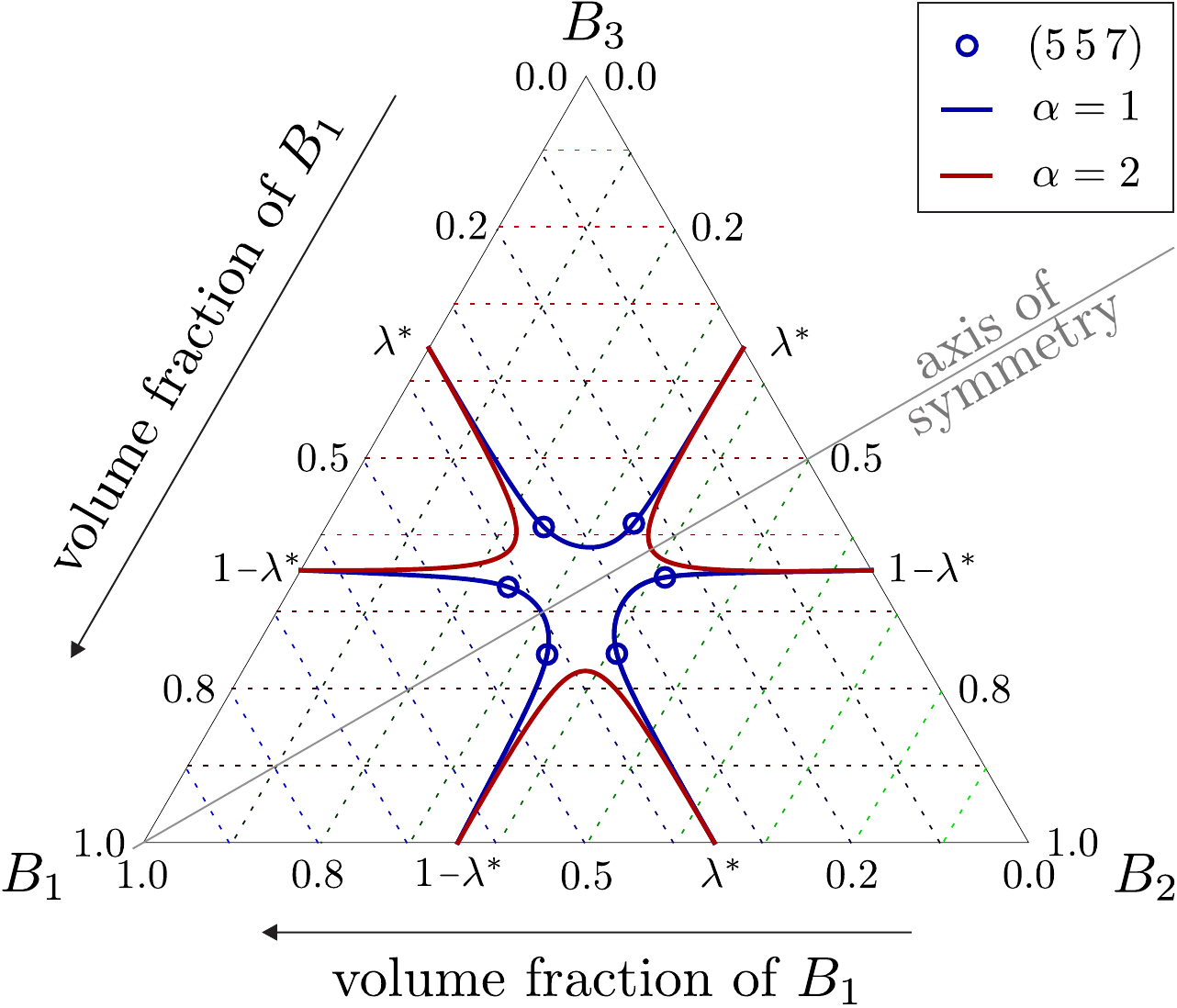}
  \caption{Ternary plot of volume fractions $1-\la$ of $B_1$, $(1-\mu^\al(\la)) \la$ of $B_2$ and $\mu^\al(\la) \la$ of $B_3$ that make the twins within twins an IPS for $\eta_1=2^{1/6}$ and $\eta_2=2^{-1/3}$.} 
  \label{FigDoubleLam}
\end{figure}

By Proposition~\ref{PropBJ}, in order to make the total shape deformation $F^\al_{\la,\m} = RB_1S^\al_2(\la,\m)S_1(\la)$ an IPS, the volume fractions $\la$, $\m$ need to be chosen such that the middle eigenvalue of $F^{\al\, T}_{\la,\m} F^\al_{\la,\m}$ is equal to one. Solving $$\det (F^{\al\, T}_{\la,\m} F^\al_{\la,\m}-\id) =0$$ for each fixed $\la$, gives rise to a quadratic equation in $\m$ for each choice of $\al$, which can be solved explicitly (see \cite{BainPaper} for the full details). The expressions are lengthy and we refer to Figure~\ref{FigDoubleLam} which visualises the dependence $\m^\al(\la)$ for the volume-preserving Bain strain. The endpoints of these curves correspond to the vanishing of one of the twinning systems (cf. \eqref{Flm}) and hence to the collapse of the system of twins within twins, to a simple twinning system with volume fractions given by \eqref{Eqmu}. The figure remains qualitatively the same for typical lattice parameters.
For each $\al$ and each admissible pair $(\la,\m^\al(\la))$ with corresponding strain $F^\al_{\la,\m^\al(\la)}$, we can calculate two one-parameter families of habit plane normals through Proposition~\ref{PropBJ}. By considering all possible combinations of twinning systems in our construction, we then obtain all crystallographically equivalent normals. For the volume-preserving Bain strain, the habit plane normals that can arise from the family $F^{1}_{\la,\m^1(\la)}$ are visualised in Figure~\ref{FigFirstSolLine}. However, due to algebraic complexity, it is difficult to write down a formula for the habit plane normals with an explicit dependence on $\eta_1$, $\eta_2$ and $\la$.
\subsection*{Why twins within twins?}
It is natural to assume that the observed total shape deformation $F$ requires small overall atomic movement (see also \cite{BainPaper}) relative to the parent phase of austenite. A measure of this distance is the \emph{strain energy}\footnote{Alternatively, one may use $\dm(F,\id) =\sum^3_{i=1}(\nu_i(F)-1)^2$ which yields the same results.} given by
\begin{equation*}
\dm(F,\id) = |F^TF-\id|^2=\sum^3_{i=1}(\nu^2_i(F)-1)^2,
\end{equation*}
where $|A|^2=\Tr(A^TA)$ denotes the Frobenius norm and $\nu_i(F)$ the principal stretches of $F$. It can be shown that any microstructure with small strain energy must necessarily involve all three Bain variants in roughly similar volume fractions. In particular, this cannot be the case for an array of twin related variants and we ought to consider at least twins within twins (see also Fig.~\ref{FigDis}). Although, introducing even further levels of twinning can reduce the strain energy, one could argue that interfacial energy contributions, which are not accounted for in this model, may inhibit such behaviour. 

\section{A new theory for the (5\,5\,7) lath transformation}
Combining the fact that $\{5\,5\,7\}$ cannot result from simple twinning (cf. Fig.~\ref{FigSimplNormals}) and that twins within twins are preferable in terms of strain energy, we build a theory that predicts $\{5\,5\,7\}$ habit plane normals solely based on energy minimisation and geometric compatibility.  

\begin{figure*}[ht]
  \centering
  \includegraphics[width=.95\columnwidth]{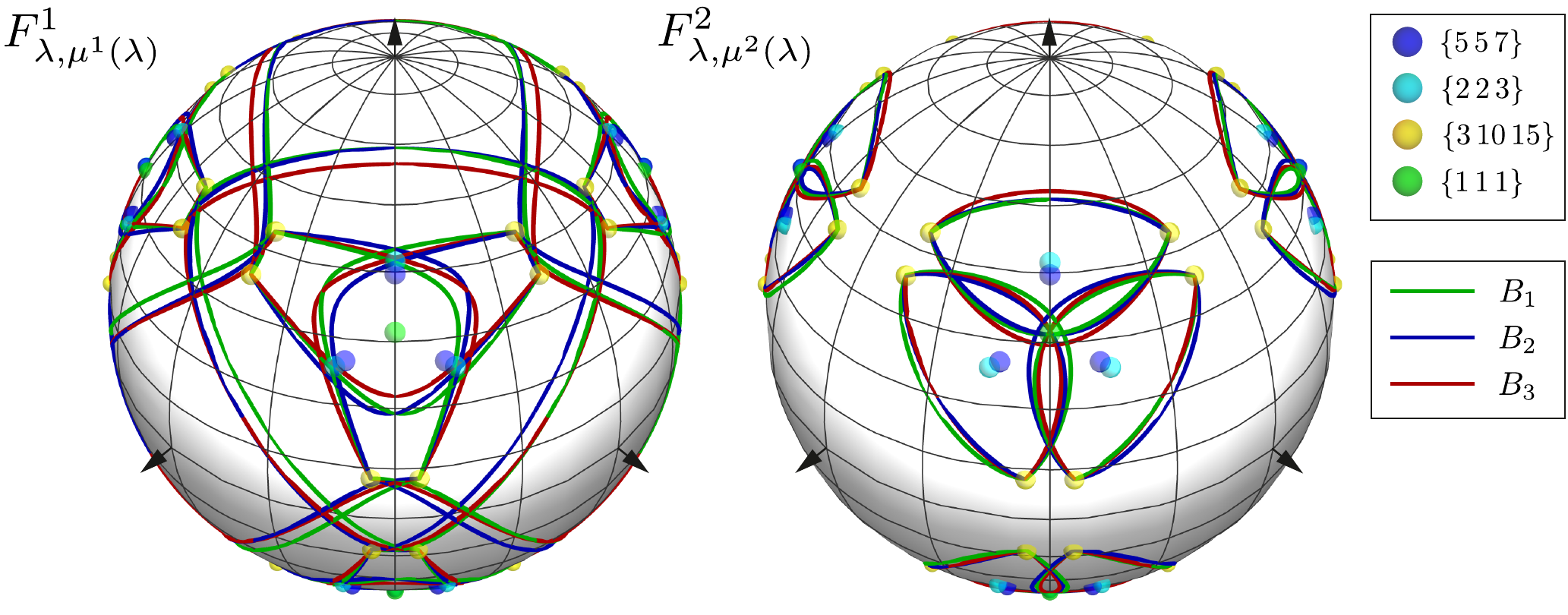}
  \caption{Possible habit plane normals for $F^{1}_{\la,\m^1(\la)}$ and $F^{2}_{\la,\m^2(\la)}$ with $\eta_1=2^{1/6}$ and $\eta_2=2^{-1/3}$. Yellow points correspond to habit plane normals arising from simple twinning.} 
  \label{FigNorm}
\end{figure*}

\begin{figure}[h]
  \centering
  \includegraphics[width=0.61\columnwidth]{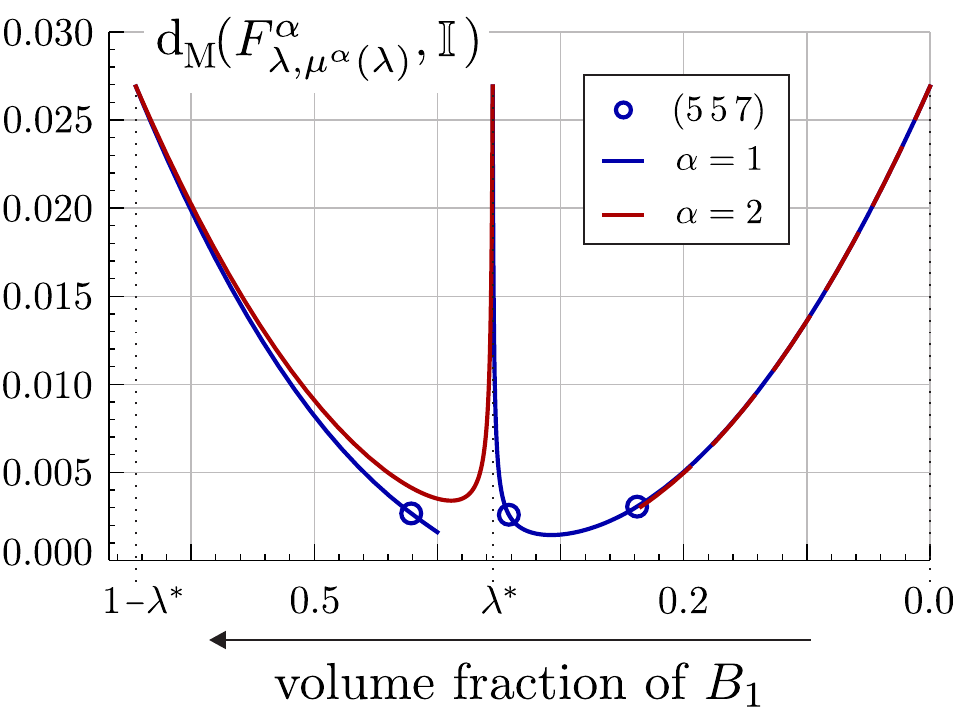}
  \caption{Strain energies for $\eta_1=2^{1/6}$ and $\eta_2=2^{-1/3}$.} \label{FigDis}
\end{figure}

Firstly, the one-parameter families of habit plane normals obtained from twins within twins (see Section~\ref{SecDoubleShear}), contain normals very close to any $\{5\,5\,7\}$. This is at least the case for lattice parameters close to $\eo=2^{1/6}\approx 1.12$,  $\et=2^{-1/3}\approx 0.79$ corresponding to a volume-preserving transformation from fcc to bcc. This regime of parameters is suitable since $\{5\,5\,7\}$ habit planes are observed in low-carbon steels where the transformation is very nearly fcc to bcc.
%For a more detailed exposure on the optimality of the Bain strain and the correct choice of lattice parameters in an fcc to bcc transformation see \comt{Bain paper}.
The resulting one-parameter families of habit plane normals, along with their crystallographically equivalent ones, are shown in Figure~\ref{FigNorm}. We stress that the only free parameter in the generation of these normals is $\la$ which fixes the choice of the shearing systems, based only on the energy minimising property of the microstructure. 
Secondly, out of these one-parameter families of normals, our theory can identify the $\{5\,5\,7\}$ habit plane normals as those satisfying a criterion of maximal compatibility. To this end, revisiting our construction of twins within twins there is a choice (cf. \eqref{Flm}) of using either $F^1_{\la,\m^1(\la)}$ as an average strain, corresponding to the Type I solution from Mallard's Law, or $F^2_{\la,\m^2(\la)}$ corresponding to Type II. Figure~\ref{FigDis} shows the strain energy associated with the two macroscopic strains as a function of $\la$. It is clear that the strain energy of $F^1_{\la,\m^1(\la)}$ is significantly smaller than that of $F^2_{\la,\m^2(\la)}$ and is thus preferable. We also note that, in agreement with the previous section, the strain energies of both $F^1$ and $F^2$ increase rapidly as the volume fraction of $B_1$ approaches $0,\, \la^*$ or $1- \la^*$, that is as the microstructure reduces to a single twinning system. Further, we remark that the  $F^2_{\la,\m^2(\la)}$ with minimal 
strain energy result in 
habit plane normals which are very nearly $\{1\,1\,1\}$ (see also Fig.~\ref{FigNorm}). Nevertheless, the strain energies of any of the $F^1_{\la,\m^1(\la)}$ that give rise to $\{5\,5\,7\}$ normals are 
lower (cf. Fig.~\ref{FigDis}).
Even though, the strains resulting in $\{5\,5\,7\}$ habit plane normals do not minimise the strain energy, they satisfy a strong criterion of compatibility. To understand this one must think in terms of the dynamic process of nucleation. As austenite is rapidly quenched, the martensite phase nucleates at various sites. The strain in a given nucleation site may need to be an IPS but otherwise, has no reason to be the same as the strain in any other site. Nevertheless, there are essentially only three distinct families of systems of twins within twins and these can be classified by the Bain variant which is present in both of the simple twinning systems that comprise the overall microstructure. In Figure~\ref{FigNorm}, these three families are distinguished by colour. As the nuclei grow and approach other nuclei, they need to remain compatible with each other. Remarkably, the only habit plane normals that arise from deformations with low strain energy and can be reached by all three families are $\{5\,5\,7\}$. 
In 
Figure~\ref{FigNorm}, this can be seen from the fact that all differently coloured curves intersect close to $\{5\,5\,7\}$. For any two such regions of twins within twins with corresponding average strains $\id+\bb_1\tp (5\,5\,7)$ and $\id+\bb_2\tp (5\,5\,7)$, one can see that
\[
(\id+\bb_1\tp (5\,5\,7)) - (\id+\bb_2\tp (5\,5\,7)) = (\bb_1 - \bb_2)\tp (5\,5\,7),
\] 
implying that they can meet along a fully coherent planar interface of normal $(5\,5\,7)$. Of course, any nucleus interacts with its neighbours faster than it does with distant nuclei. As a result, blocks of similarly oriented regions of twins within twins (laths) may form whose overall orientation may differ from that of other blocks. 
\begin{figure*}[ht]
  \centering
  \includegraphics[width=\columnwidth]{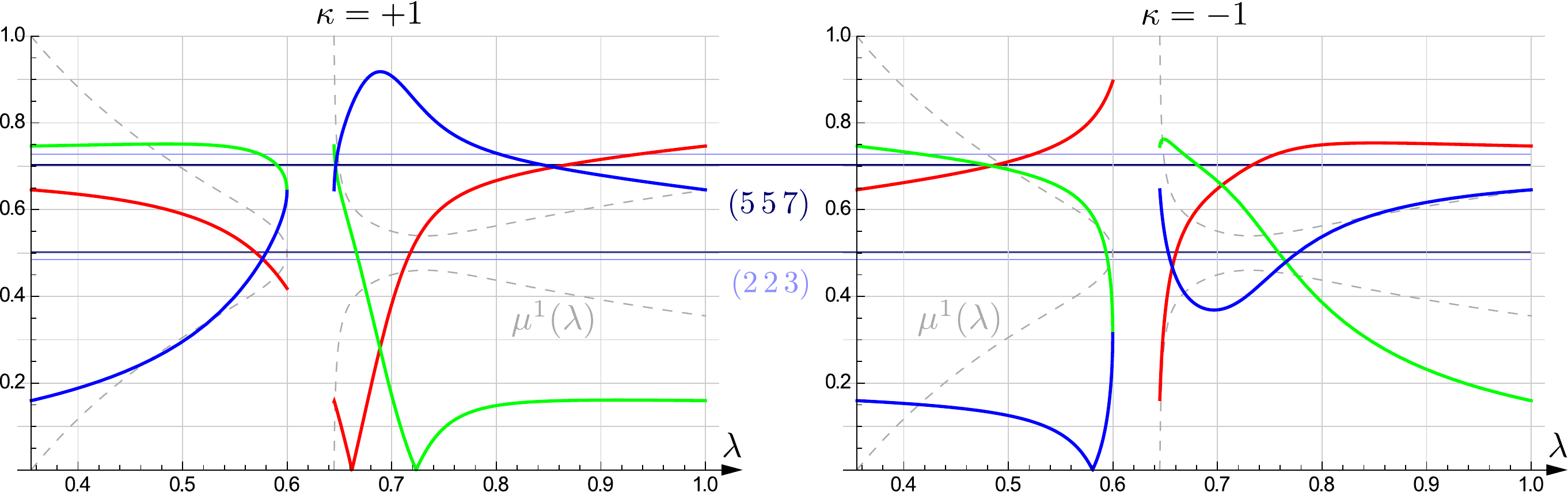}
  \caption{Coordinates of the habit plane normals (blue,\,red,\,green) for the macroscopic strains $F^{1}_{\la,\m^1(\la)}$ and $\kappa \in \{-1,1\}$ (cf. Proposition~\ref{PropBJ}). The lattice parameters are $\eta_1=2^{1/6}$ and $\eta_2=2^{-1/3}$, corresponding to the volume-preserving Bain strain.} 
  \label{FigFirstSolLine}
\end{figure*}
\section{Concluding remarks} \label{SecConcl}
The theory proposed here for the prediction of $(5\,5\,7)$ habit plane normals has two possible interpretations. On the one hand,
it can be seen as a purely macroscopic theory. In particular, it is an instance of a double shear theory with a precise algorithm to produce the required shears based on energy minimisation, without the need for any further assumptions. All possible habit plane normals that can arise from the one parameter family (indexed by $\la$) of macroscopic deformations $F^{1}_{\la,\m^1(\la)}$ are shown in Figure~\ref{FigFirstSolLine}. Table~\ref{TableShear} then lists the elements of the twinning systems for the values of $\la$ that produce a near $(5\,5\,7)$ habit plane. With the help of \eqref{Flm} it is easy to convert between the twinning and shearing systems and thus compute the elements $S_1$ and $S_2$ required in a double shear theory. At this macroscopic level it is not possible to distinguish between twins within twins, a single twin and one slip system, and a single variant and two slip systems. 

On the other hand, a physical mechanism for the formation of $(5\,5\,7)$ habit plane normals is proposed and thus a specific morphology on a microscopic level. According to this interpretation, each lath may itself be a region of twins within twins with a corresponding lath boundary of normal $(5\,5\,7)$. This type of morphology is depicted in Figure~\ref{FigDouble} with $\mathbf{k}$ being a $\{5\,5\,7\}$ normal and the other elements are as in Table~\ref{TableShear}. This morphology is a direct consequence of the underlying theory and it would be very interesting if it could be put to experimental scrutiny. 
%
%Another interpretation is that due to the small length scale of the laths and the large number of interfaces the yield point is reached and as discussed in Section~\ref{SecModel}, plasticity will then only change the internal structure of the lath while maintaining its macroscopic strain. In this case the internal structure of the laths may be described by either a single variant and a system of two slips or a twin and one slip system (see \cite{BainPaper} for details). Any of these morphologies is a direct consequence of the underlying theory and it would be very interesting if it could be put to experimental scrutiny.
%
\begin{table}[ht]
\begin{center}
  \begin{tabular}{l*{4}{c}}
$\la$&$0.576$ & $0.659$& $0.762$\\
\hline
$\aaa$&$[.374, -.529, 0]$ & $[.374, -.529, 0]$& $[.374, -.529, 0]$\\
$\nn$&$2^{-\frac{1}{2}}(1, 1, 0)$& $2^{-\frac{1}{2}}(1, 1, 0)$& $2^{-\frac{1}{2}}(1, 1, 0)$\\
$\mu^1(\la)\mkern-18mu$&$0.581$ & $0.621$ & $0.546$\\
$\bb_\la$& $[.130, .234, .315]$ & $[.135, .260, .359]$& $[.137, .289, .412]$\\
$\mm$ & $2^{-\frac{1}{2}}(0,1,-1)$ & $2^{-\frac{1}{2}}(0,1,-1)$ & $2^{-\frac{1}{2}}(0,1,-1)$
\end{tabular}
\caption{Elements of the twinning system \eqref{Flm} leading to $\{5\,5\,7\}$ habit plane normals. The remaining $\{5\,5\,7\}$ normals can be obtained from the crystallographically equivalent systems.} \label{TableShear}
\end{center}
\end{table}
\section*{Acknowledgements}
The research of A. M. leading to these results has received funding from the European Research Council under the European Union's Seventh Framework Programme (FP7/2007-2013) / ERC grant agreement n$^\circ\, 291053$.

\newcommand{\etalchar}[1]{$^{#1}$}

%\bibliography{Bib}
%\bibliographystyle{unsrt}
%\bibliographystyle{alpha}

\end{document}